\documentclass[journal]{IEEEtran}
\IEEEoverridecommandlockouts
\usepackage{cite}
\usepackage{amsmath,amssymb,amsfonts}
\usepackage[colorlinks=true,linkcolor=blue]{hyperref}
\usepackage{algorithmic}
\usepackage{graphicx}
\usepackage{textcomp}
\usepackage{xcolor}
\usepackage{floatrow}
\usepackage{caption}
\usepackage{subcaption}

\def\BibTeX{{\rm B\kern-.05em{\sc i\kern-.025em b}\kern-.08em
    T\kern-.1667em\lower.7ex\hbox{E}\kern-.125emX}}
\begin{document}

\title{Complex Network Analysis of the Bitcoin Transaction Network}

\author{Bishenghui~Tao,
Hong-Ning~Dai,
~Jiajing~Wu,
~Ivan~Wang-Hei~Ho,
~Zibin~Zheng,
and Chak Fong Cheang

\thanks{This work was supported in part by the Natural Science Foundation of China under Grant No. 61973325, by Guangdong Basic and Applied Basic Research Foundation under Grant No. 2021A1515011661, by Hong Kong Institute of Business Studies (HKIBS) Research Seed Fund with Grant No. HKIBS RSF-212-004. (Corresponding author: Chak Fong Cheang.)}
\thanks{B. Tao and C. F. Cheang are with Macau University of Science and Technology, Macau (email: bishenghui.tao@connect.polyu.hk, cfcheang@must.edu.mo).}
\thanks{H.-N. Dai is with Lingnan University, Hong Kong (email: hndai@ieee.org).}
\thanks{J. Wu and Z. Zheng are Sun Yat-sen University, Guangzhou, China (email: wujiajing@mail.sysu.edu.cn and zhzibin@mail.sysu.edu.cn).}
\thanks{I. W.-H. Ho is with The Hong Kong Polytechnic University, Hong Kong (email: ivanwh.ho@polyu.edu.hk).}%

}

\maketitle
\thispagestyle{plain}
\pagestyle{plain}



\begin{abstract}
In this brief, we conduct a complex-network analysis of the Bitcoin transaction network. In particular, we design a new sampling method, namely random walk with flying-back (RWFB), to conduct effective data sampling. We then conduct a comprehensive analysis of the Bitcoin network in terms of the degree distribution, clustering coefficient, the shortest-path length, connected component, centrality, assortativity, and the rich-club coefficient. We obtain several important observations including the small-world phenomenon, multi-center status, preferential attachment, and non-rich-club effect of the current network. This work brings up an in-depth understanding of the current Bitcoin blockchain network and offers implications for future directions in malicious activity and fraud detection in cryptocurrency blockchain networks.
\end{abstract}


\section{Introduction}
\IEEEPARstart{I}{n} recent years, Bitcoin has been widely accepted as one of the most representative cryptocurrencies since its genesis in 2009 by Satoshi Nakamoto~\cite{nakamoto2019bitcoin}. Owing to its innovative characteristics such as anonymity, low transaction fee, decentralization, and nonstop service accessibility, Bitcoin received extensive attention in the past several years~\cite{8653347,10.1145/3366423.3380103,9097307,9295385}. However, there are few studies on the analysis of the Bitcoin network. It is crucial to analyze the Bitcoin blockchain from the complex network perspective since it can unravel incumbent blockchain systems' enigmas. There are several attempts in this field. Refs.~\cite{baumann2014exploring} and~\cite{lischke2016analyzing} initially explore the Bitcoin transaction network with a preliminary graph analysis while~\cite{nerurkar2020dissecting} provides the longitudinal network-based analysis of the Bitcoin systems. The study~\cite{kondor2014rich} analyzes the structure of the transaction network by measuring network characteristics, including the degree distribution, degree correlations, and clustering. 
 
Nevertheless, the research on the Bitcoin transaction network is still quite limited. Existing Bitcoin network research primarily focuses on large-scale and long-term data analysis of the entire network~\cite{lischke2016analyzing}~\cite{nerurkar2020dissecting}, or transaction pattern detection~\cite{9332279}, but lacks both in-depth exploration on network characteristics and computationally-efficient algorithms of networks analysis. For example, the Bitcoin transaction network's characteristics such as assortativity, connection tendency, and centrality need to be further investigated by a computationally-efficient network analysis method. To this end,  we conduct a multi-dimensional analysis of the Bitcoin transaction network from different perspectives, including the degree distribution, small-world determination, connected component, centrality, assortativity analysis, and rich-club coefficient. The main contributions of this brief are summarized as follows.

\begin{itemize}
    \item We design a new sampling method called random walk with flying-back (RWFB) for the Bitcoin transaction data analysis. Our RWFB better preserves the characteristics of the original Bitcoin network while having less computational complexity than conventional methods (\S~\ref{sec:sampling}).
    \item We conduct a multi-dimensional analysis of the Bitcoin transaction network, from perspectives of the degree distribution, clustering coefficient, shortest-path length, connected component, centrality, assortativity, and rich-club coefficient (\S~\ref{sec:analysis}). 
    \item We obtain several novel observations of the current Bitcoin network, such as the small-world effect, one-way transactions, multi-center phenomenon, robustness, disassortativity, and non-rich-club ordering (\S~\ref{sec:analysis}).
\end{itemize}

\section{Network construction}
When users transfer Bitcoin (BTC) from one address to another, transactions are made. In Bitcoin, one transaction may have multiple input addresses and multiple output addresses. We extract a transaction having inputs from $x$ addresses and outputs to $y$ addresses, thereby being represented by $x\times y$ edges~\cite{kondor2014rich}. Thereafter, each node represents a Bitcoin address, every directed edge represents the flow of funds of the transaction, and the weight of the edge is proportional to the transaction value. Consider an example, in which a transaction has two inputs $A$ and $B$ paying 2 and 8 BTC, respectively, and three outputs $C$, $D$, and $E$ receiving 2, 3, and 4 BTC, respectively (note that the sum of outputs is not equal to that of inputs due to 1 BTC for the transaction fee). Thus, the weight of the edge from $A$ to $D$ is $2/(2+8)\times3=0.6$.

We then convert a Bitcoin transaction network into a weighted directed graph denoted by $G=(V,E,W)$, where $V$ is a set of nodes, $E$ is a set of edges, and $W$ is a set of weights. Each edge is represented as $e_{ij}=(i,j,w_{ij})$, where $i$ denotes the input node, $j$ denotes the output node, and $w_{ij}$ denotes the weight value. The set $E$ with $N$ nodes can be represented by an $N \times N$ matrix, which is essentially an adjacency matrix denoted by $\mathbf{A}$. For any element $a_{ij}$ in $\mathbf{A}$, we have $a_{ij}=w_{ij}$ if there exists a link with weight $w_{ij}$ between $i$ and $j$; $a_{ij}=0$ otherwise~\cite{chen2014fundamentals}. In particular, we have
\begin{equation}
\footnotesize
a_{ij}=
\left\{\begin{matrix}
w_{ij}         &  \text { if }\ e _{ij}\ \text { is\ defined;}\\ 
0 & \ \ \ \ \  \text { if } \ e _{ij} \ \text { is\  not\ defined}.
\end{matrix}\right.
\label{equ:aj}
\end{equation}

\begin{figure}
\floatsetup{captionskip=-0.001cm}
\ffigbox[8.9cm]{%
\begin{subfloatrow}
  \ffigbox[\FBwidth][]
    {\caption{Visualization of the Bitcoin network.}\label{fig:sampled}}
    {\includegraphics[width=4.2cm]{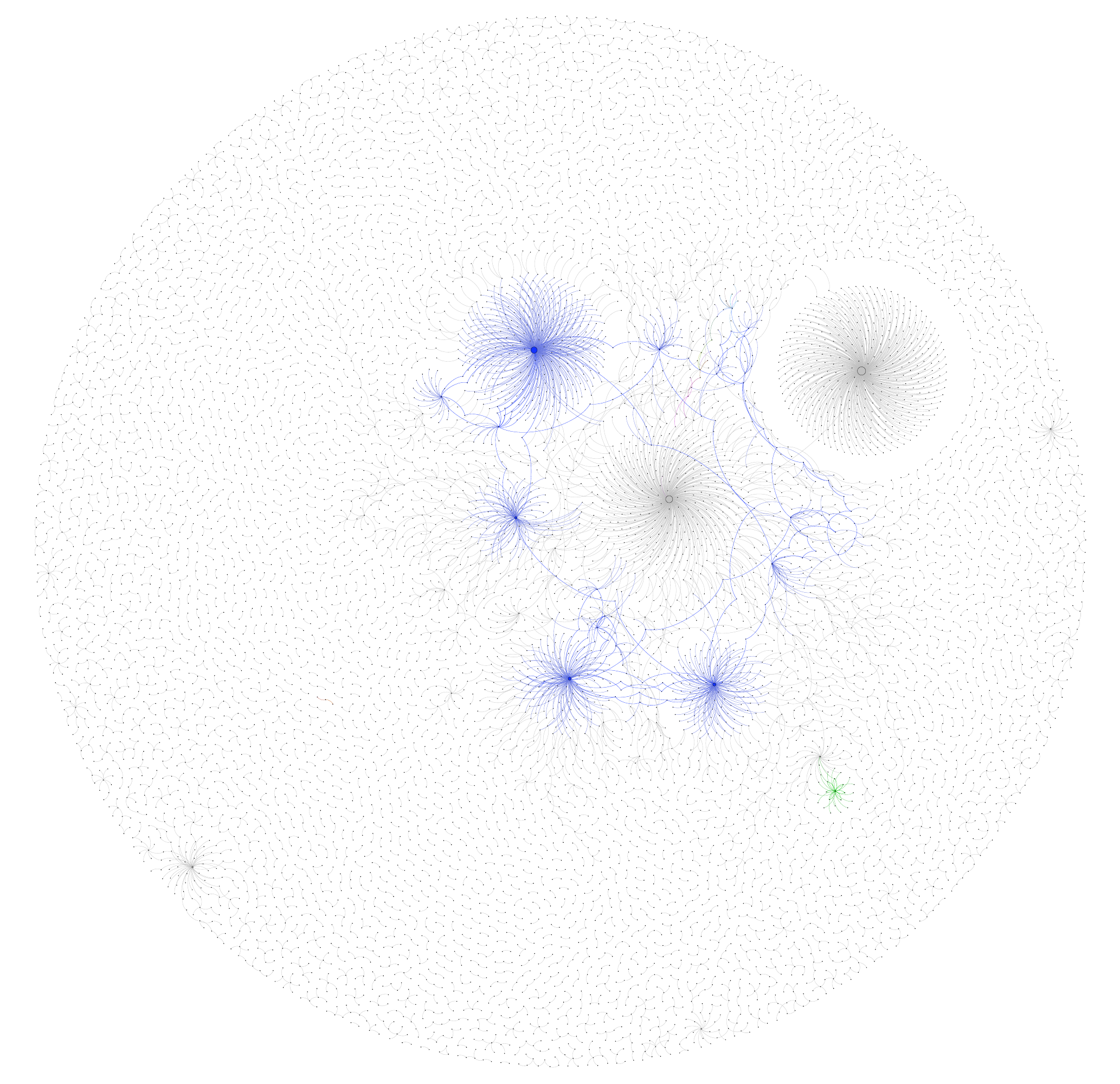}}
\end{subfloatrow}
\hspace*{\columnsep}
\begin{subfloatrow}
  \hsize0.1\hsize
   \vbox to 4.15cm{
  \ffigbox[\FBwidth]
    {\caption{Comparison of different sampling methods on graph kernel.}\label{fig:gk}}
    {\includegraphics[width=4.0cm]{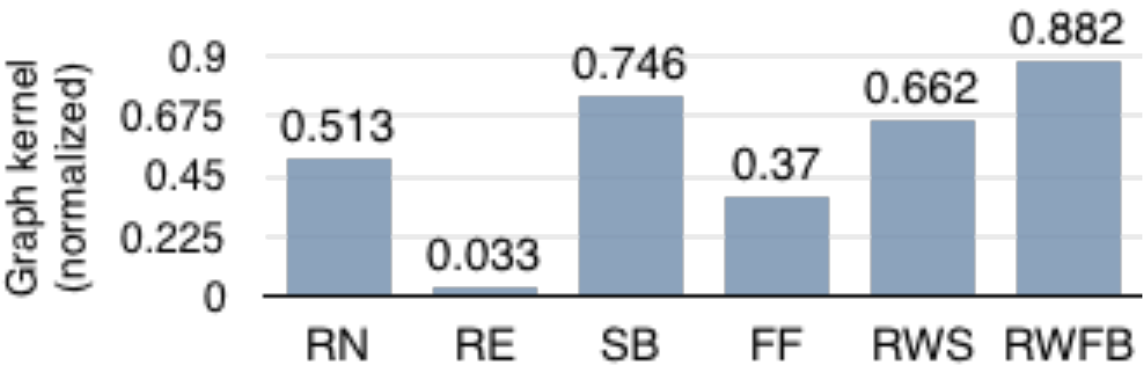}}
    \vss
  \ffigbox[\FBwidth]
    {\caption{Parameter sensitivity of parameter $p$}\label{fig:p}}
    {\includegraphics[width=4.0cm]{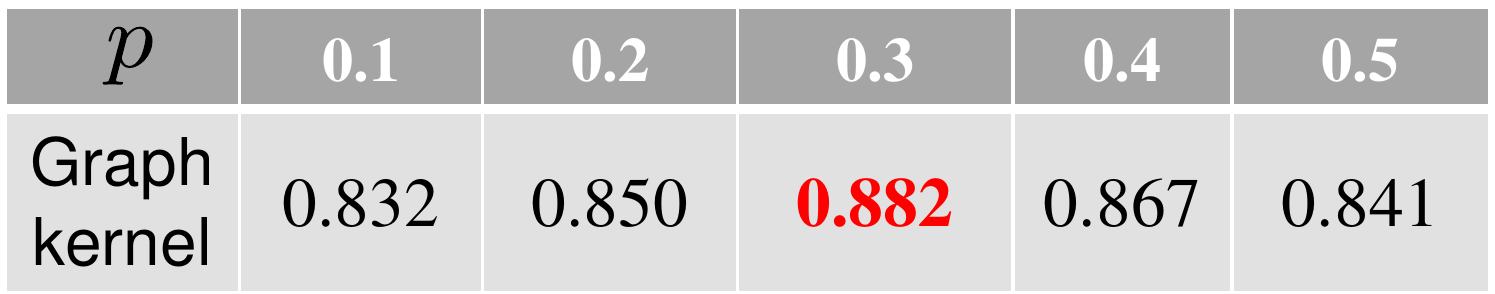}}
   }
\end{subfloatrow}
}
{
\caption{Graph sampling results}\label{fig:system}}
\vspace*{-0.5cm}
\end{figure}

We obtain the transaction and address data by the Bitcoin Project database~\cite{kondor2014inferring} from January 2017 to January 2018 via synchronizing real-time Bitcoin blockchain. The data we considered includes 148,803,071 nodes and 871,517,687 edges. Fig.~\ref{fig:system}(\subref{fig:sampled}) presents a graph view of the Bitcoin blockchain transaction network with 10,000 selected nodes.

\section{Bitcoin network sampling}
\label{sec:sampling}


\subsection{Random-walk based graph sampling}

Since the Bitcoin transaction network is a large massive graph with millions of nodes, it is necessary to obtain a representative sample network for simplifying the analysis. Some pioneering studies~\cite{lin2020modeling} and~\cite{perozzi2014deepwalk} show that random walk-based sampling methods can well preserve structural properties of the scale-free blockchain network with a power-law degree distribution. Therefore, we also design a graph sampling method based on random walk to represent the Bitcoin blockchain network.

In traditional random-walk methods like~\cite{spitzer2013principles}, the next-hop node $j$ is randomly chosen from the neighbors of the current node $i$. However, these methods cannot accurately sample the Bitcoin network since they only choose one neighbor in every step, consequently leading to inaccurate graph properties, e.g., a lower average degree~\cite{yoon2007statistical}. To address this problem, we propose an improved sampling method, namely RWFB. In particular, RWFB considers a flying-back probability when sampling the network. At every step of the directed random walk, RWFB flies back to the current node $i$ with the flying-back probability $p$; it chooses a random neighbor among its $k_{i}$ neighbors to move to with a probability of $1-p$, thereby each having a probability of $\frac{(1-p)}{k_{i}}$. Therefore, we have the RWFB probability denoted by $\mathbb{P}_{i}^{\textup{RWFB}}$ defined as follows
\begin{equation} 
\footnotesize
\mathbb{P}_{i}^{\text{RWFB}}=\left\{\begin{array}{ll}
\frac{(1-p)}{k_{i}} & \text  { move to neighbor\ } j \text { of node } i, \\
p & \text { fly back to node\ } i.
\end{array}\right.
\end{equation}
Thereafter, we redefine the sampled graph with flying-back function as $G_{\textup{RWFB}}=(V_{i},E_{i,j},w_{i,j})$. It is a directed weighted graph, whereas the nodes are the vertices that walk through and the edges are the steps. It is worth mentioning that the walk always starts from a random node. Moreover, if it goes into an impasse (or a deadlock) during the walk, another random node will be chosen to continue until the sampling size satisfying the given value.

\subsection{Evaluation of sampling methods}

\begin{table}[t]
\floatsetup{captionskip=-0.001cm}
\centering
\caption{Comparison of sampling methods by K-S $D$-statistic}
\renewcommand{\arraystretch}{1.05}
\scriptsize
\begin{tabular}{|l|l|l|l|l|l|}
\hline
      & \multicolumn{1}{c|}{Degree} & \multicolumn{1}{c|}{Clustering} & \multicolumn{1}{c|}{Betweenness} & \multicolumn{1}{c|}{Closeness} & \multicolumn{1}{c|}{AVG} \\ \hline\hline
RWFB & \textbf{0.120}            & \textbf{0.045}                & 0.091              & \textbf{0.429}             & \textbf{0.171}           \\ \hline
RWS   & 0.293                    & 0.046                        & 0.536                        & 0.618                      & 0.373                 \\ \hline
RN    & 0.895                      & 0.053                        & 0.151                        & \textbf{0.433}                      & 0.383                 \\ \hline
RE    & 0.275                    & 1                               & \textbf{0.067}              & 0.549                     & 0.473                 \\ \hline
FF  & \textbf{0.187}                    & 1                        & \textbf{0.075}                       & 0.669                      & 0.483                 \\ \hline
SB  & 0.409                    & \textbf{0.025}                        & 0.583                       & 0.645                      & 0.415                 \\ \hline
\end{tabular}
\label{tab:results}
\vspace*{-0.5cm}
\end{table}
 
We evaluate the Bitcoin transaction graph by comparing the proposed RWFB method with other five classic sampling methods: Random Walk Selection (RWS), Random Node (RN), Random Edge (RE), Forest Fire (FF)~\cite{leskovec2005graphs}, and Snow Ball (SB)~\cite{leskovec2006sampling}. Note that the RN method randomly selects nodes while RE uniformly selects edges at random.

We evaluate its distribution of \emph{degree}, \emph{clustering coefficient}, \emph{betweenness}, and \emph{closeness}~\cite{chen2014fundamentals} for each method. The complete results are compared with the original graph by the Kolmogorov-Smirnov (K-S) $D$-statistic~\cite{leskovec2006sampling}, which measures the agreement between two distributions. The complete results and an average $D$-statistic value (AVG) of all sampling methods are presented in Table~\ref{tab:results}. In addition, a value close to zero means that the gap between the two distributions is small, implying the satisfactory sampling effect. The well-performed score in each column is highlighted in bold font. We observe that the proposed RWFB outperforms others in most of the given metrics, implying that the sampled graph obtained by our RWFB has the best approximation to the original graph. 

Moreover, we also consider the graph kernel method~\cite{T2003On}, which is an effective measurement of similarity of graphs. In particular, we adapt the shortest-path graph kernel as $K_{\text{sp}}$, to indicate the kernel of two graphs $G_{i}$ and $G_{j}$~\cite{2005Shortest}:
\begin{equation}
\setlength{\abovedisplayskip}{3pt}
\footnotesize
K_{\text{sp}}(G_{i}, G_{j})=\sum_{(u, v) \in V(G_{i})^{2} \atop u \neq v} \sum_{(w, z) \in V(G_{j})^{2}\atop w \neq z} K\big((u, v),(w, z)\big),
\setlength{\belowdisplayskip}{3pt}
\end{equation}
where $K\big((u, v),(w, z)\big)=K_{\text{D}}\big(k(u), k(w)\big) \cdot K_{\text{D}}\big(k(v), k(z)\big) \cdot K_{\text{L}}(d(u, v), d(w, z))$. The term $K_{\text{D}}$ is a kernel for comparing vertex degree and $K_{\text{L}}$ is another kernel to compare the shortest-path length. Meanwhile, $d(u, v)$ denotes the shortest-path distance between the vertices $u$ and $v$.

Fig.~\ref{fig:system}(\subref{fig:gk}) presents the results of the aforementioned sampling methods w.r.t. the normalized shortest-path graph kernel. Note that for a graph that is completely similar to itself, its normalized graph kernel value should be 1 while the kernel of a completely dissimilar graph should be 0. We select the best performing parameters and results for each method.  It is worth mentioning that we investigate the effect of parameter $p$ on the graph kernel measurement, as shown in Fig.~\ref{fig:system}(\subref{fig:p}), where the best value (0.882) is obtained when $p=0.3$. As shown in Fig.~\ref{fig:system}(\subref{fig:gk}), we observe that our RWFB reaches the highest value (0.882) closest to 1 among all methods. This result demonstrates that the graph sampled by RWFB is the most approximate to the original one, implying its superior performance in restoring the network structure than other methods. Therefore, we employ the RWFB method to sample the Bitcoin transaction network for exploring more insights in the following sections.

\section{Complex network analysis}
\label{sec:analysis}

\subsection{Degree distribution}\label{Degree}

It is crucial to investigate the node degree distribution of the Bitcoin network.  The number of adjacent edges of a node is defined as \emph{degree} denoted by $k$ in the complex network. In the Bitcoin transaction network, the degree $k$ is calculated for every Bitcoin address by summation of the number of transactions. Meanwhile, the number of incoming transactions (receiving BTC) is the \emph{in-degree} and the number of outgoing transactions (paying BTC) is the \emph{out-degree}. Moreover, we also introduce the degree distribution denoted by $P(k)$ on degree $k$. The degree distribution $P(k)$ is the probability that a randomly-selected node has the degree equal to $k$~\cite{chen2014fundamentals}. Further, if the degree $k$ obeys the power law, we then have $
P(k) \propto k^{-\alpha}$, where $\alpha$ is the scaling parameter of the power-law distribution.

\begin{figure}
\floatsetup{captionskip=-0.001cm}
\ffigbox[8.9cm]{%
\begin{subfloatrow}
  \ffigbox[\FBwidth][]
    {\caption{Degree distribution}\label{fig:a}}
    {\includegraphics[width=2.8cm]{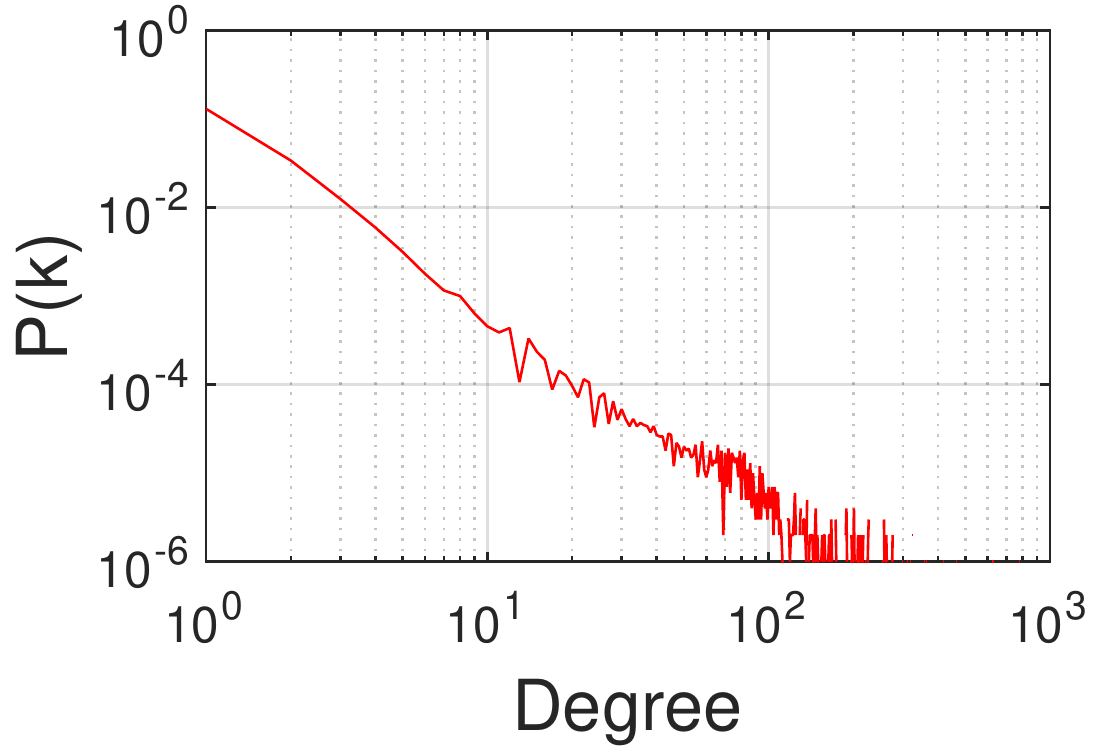}}
\end{subfloatrow}
\begin{subfloatrow}
  \ffigbox[\FBwidth][]
    {\caption{In-degree distribution}\label{fig:b}}
    {\includegraphics[width=2.8cm]{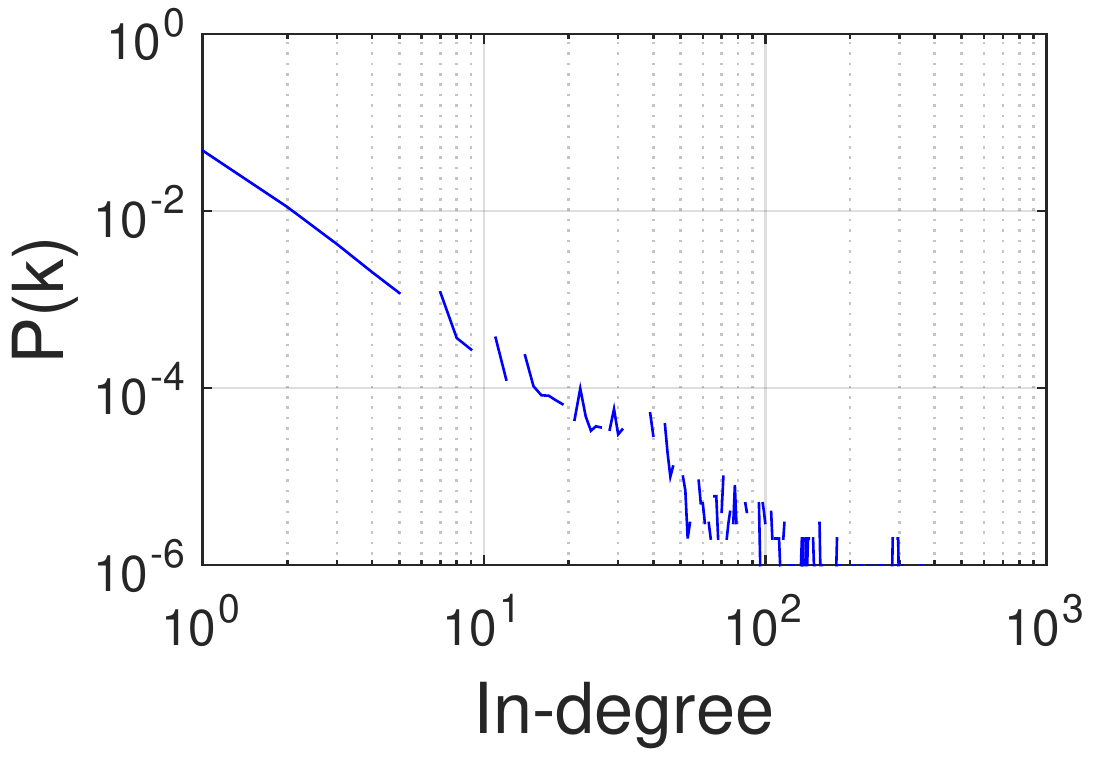}}
\end{subfloatrow}
\begin{subfloatrow}
  \ffigbox[\FBwidth][]
    {\caption{Out-degree distribution}\label{fig:c}}
    {\includegraphics[width=2.8cm]{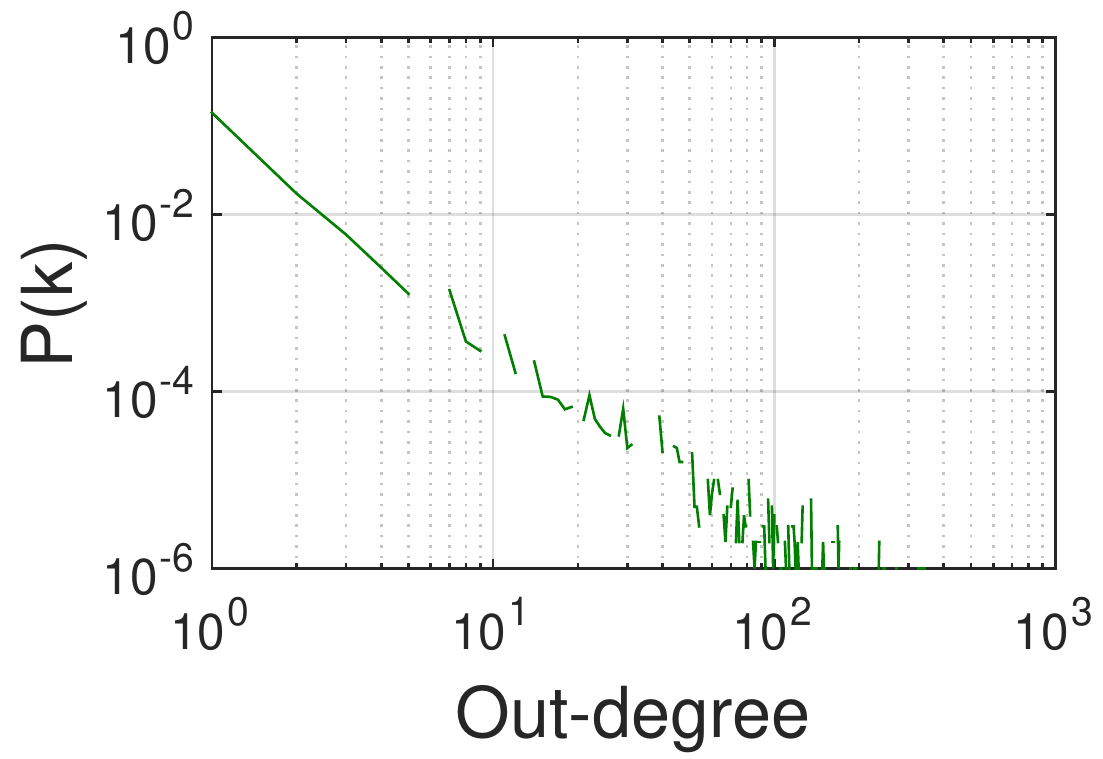}}
\end{subfloatrow}
}
{\caption{Node degree distribution of the Bitcoin network.}\label{fig:degree}}
\vspace*{-0.5cm}
\end{figure}

Fig.~\ref{fig:degree} shows the degree distributions of the Bitcoin blockchain network in log-log plots (i.e., logarithmic scale in both horizontal and vertical axes). In particular, the scaling parameter of the total degree distribution is $\alpha=1.411$ as shown in Fig.~\ref{fig:degree}(\subref{fig:a}). Furthermore, we also consider the in-degree in Fig.~\ref{fig:degree}(\subref{fig:b}) (i.e., $\alpha_{\text {in}}=1.421$) and out-degree distributions Fig.~\ref{fig:degree}(\subref{fig:c}) (i.e., $\alpha_{\text {out}}=1.418$), respectively.  
We observe from the results that all degree distributions follow the power-law distribution with the heavy-tail. It implies that the Bitcoin network is a scale-free network, in which only a few nodes have a large number of connections~\cite{clauset2009power} while the majority of nodes are of low degrees and have fewer connections. This is consistent with the findings obtained by the research~\cite{nerurkar2020dissecting, lischke2016analyzing} using complete real network data.

\subsection{Clustering coefficient and the shortest-path length}\label{Clustering}

The clustering coefficient and the shortest-path length can measure the network from a geometric point of view.
We denote the average clustering coefficient of the network by $C$, which is given as follows
\begin{equation}
\setlength{\abovedisplayskip}{3pt}
\footnotesize
	C=\frac{1}{N}\sum_{i\in V(G)}\frac{\left | \Delta _{i} \right |}{k_{i}(k_{i}-1)/2},
\label{equ:c}
\setlength{\belowdisplayskip}{3pt}
\end{equation}
where $N$ denotes the number of nodes, $\left | \Delta _{i} \right |$ is the number of complete triangles and $k_{i}$ denotes the degree of node $i$.

On the other hand, we denote the average shortest path length by $L$, which is expressed by 
\begin{equation}
\setlength{\abovedisplayskip}{3pt}
\footnotesize
L=\sum_{i, j \in V(G)} \frac{l(i, j)}{N(N-1)},
\label{equ:L}
\setlength{\belowdisplayskip}{3pt}
\end{equation}
where $V(G)$ is the set of nodes in graph $G$ and $l(i, j)$ denotes the shortest path length from $i$ to $j$. For the Bitcoin transaction graph, we have $C_{\text{Bitcoin}}=0.0071$ and $L_{\text{Bitcoin}}=3.833$, implying that there are many indirect transactions~\cite{chen2014fundamentals}. 

\textbf{Small-world effect of the Bitcoin network}. We observe that the Bitcoin network conforms to the small-world network model according to $C_{\text{Bitcoin}}$ and $L_{\text{Bitcoin}}$. To find the evidence of this observation, it is necessary to compare the Bitcoin network with the equivalent random graph and the lattice graph~\cite{telesford2011ubiquity}. Thus, we denote the average shortest-path length of the random network by $L_{\text{rand}}$, and the average clustering coefficient of the lattice network by $C_{\text{latt}}$. The small-world measurement denoted by $\omega$~\cite{watts1998collective} is then defined as follows
\begin{equation}
\setlength{\abovedisplayskip}{3pt}
\footnotesize
\omega=\frac{L_{\text {rand }}}{L_{\text{Bitcoin}}}-\frac{C_{\text{Bitcoin}}}{C_{\text {latt }}}, 
\setlength{\belowdisplayskip}{3pt}
\end{equation} 
where we consider several different random networks and lattice networks in each result. In this brief, we count five generated random networks and lattice networks to compute the mean value of $L_{\text{rand}}$ and $C_{\text{latt}}$, respectively. The value of $\omega$ is restricted to the range $[-1,1]$. The network is considered as a small-world network whereas $\omega$ is close to zero. Table~\ref{tab:results2} shows that $\omega= 0.23$, indicating that the Bitcoin network is indeed a small-world network according to the characteristics as in~\cite{watts1998collective}. This effect implies that Bitcoin tokens can be transferred among the majority by a few steps. 

\begin{table}[t]
\setlength{\abovecaptionskip}{-0.02cm}
\centering
\caption{Small-world metrics of the graph}
\scriptsize
\begin{tabular}{|l|l|l|l|l|}
\hline
$L_{\text{Bitcoin}}$ & $L_{\text{rand}}$ & $C_{\text{Bitcoin}}$ & $C_{\text{latt}}$ & $\omega$ \\ \hline
3.833                & 9.00              & 0.0071              & 0.0033            & 0.23     \\ \hline
\end{tabular}
\vspace*{-0.5cm}
\label{tab:results2}
\end{table}

\subsection{Connected component}\label{Connected component}

Since we have found that the Bitcoin blockchain network is a small-world network, analyzing its connectivity is of great significance. In a complex network, if any pair of nodes in a sub-graph has at least one connected path, we call this sub-graph a \emph{connected component}. Meanwhile, in the case of a directed network, we measure its \emph{strongly-connected components} (SCCs), in which any random node pair $(i,j)$ has a directed path from $i$ to $j$ and a directed path from $j$ to $i$ simultaneously. Similarly, \emph{weakly-connected components} (WCCs) refer to undirected connected components. 

Table~\ref{tab:cc} presents the calculation results of connected components, including the number of SCCs, the size of the largest SCC, the number of WCCs, and the size of the largest WCC. We observe that both the largest SCC and the largest WCC are relatively sizable compared to the entire graph's size (covering about 45$\%$ and 93$\%$ of the nodes, respectively), implying a relatively-connected graph. We also speculate that existing hub nodes bridge many isolated nodes. In reality, such a hub node may be an exchange, a trading institution, or a financial organization. Meanwhile, it can be witnessed that the number of WCCs is far smaller than that of SCCs, indicating that many transactions are only one-way in this graph. In other words, most nodes do not make bidirectional deals (input and output) frequently, i.e., they only pay or only accept BTC.

\begin{table}[h]
\centering
\caption{Connected component metrics of the graph}
\renewcommand{\arraystretch}{1.15}
\scriptsize
\begin{tabular}{|l|l|l|l|l|}
\hline
\#Nodes & \#SCC  & Largest SCC & \#WCC & Largest WCC \\ \hline
1,000,000 & 548,739 & 450,442                                                & 13,638 & 930,621                                                \\ \hline
\end{tabular}
\label{tab:cc}
\vspace*{-0.5cm}
\end{table}

\subsection{Centrality}\label{Centrality}

The analysis of the connected components leads to our conjecture about the existence of hub nodes in the Bitcoin network. To verify that, we then further analyze its centrality, which measures the relative structure importance of the nodes in the graph. Firstly, in the Bitcoin network, the \emph{closeness centrality} is a metric to measure how long it will take when a node transforms the fund to all the others sequentially. The closeness centrality denoted by $O_{(i)}$ of node $i$ is defined as
\begin{equation}
\setlength{\abovedisplayskip}{3pt}
\footnotesize
O_{(i)}=\frac{n-1}{\sum_{j=1}^{n-1} d(i, j)},
\setlength{\belowdisplayskip}{3pt}
\end{equation}
where $n$ is the number of node $i$'s reachable nodes and $d(j, i)$ is the shortest-path distance between node $j$ and node $i$. Moreover, since we already confirm that this graph has more than one connected component, we introduce the improved closeness-centrality measurement proposed by Wasserman and Faust~\cite{wasserman1994social} denoted by $O_{\text{WF}}(i)$ as follows,
\begin{equation}
\setlength{\abovedisplayskip}{3pt}
\footnotesize
O_{\text{WF}}(i)=\frac{(n-1)^{2}}{(N-1) \sum_{j=1}^{n-1} d(j, i)},
\setlength{\belowdisplayskip}{3pt}
\end{equation}
where $N$ denotes the total number of nodes in the graph.

\begin{figure}
\floatsetup{captionskip=-0.001cm}
\ffigbox[8.9cm]{%
\begin{subfloatrow}
  \ffigbox[\FBwidth][]
   {\caption{Closeness and betweenness versus node degree.}\label{fig:centrality}}
    {\includegraphics[width=4.9cm]{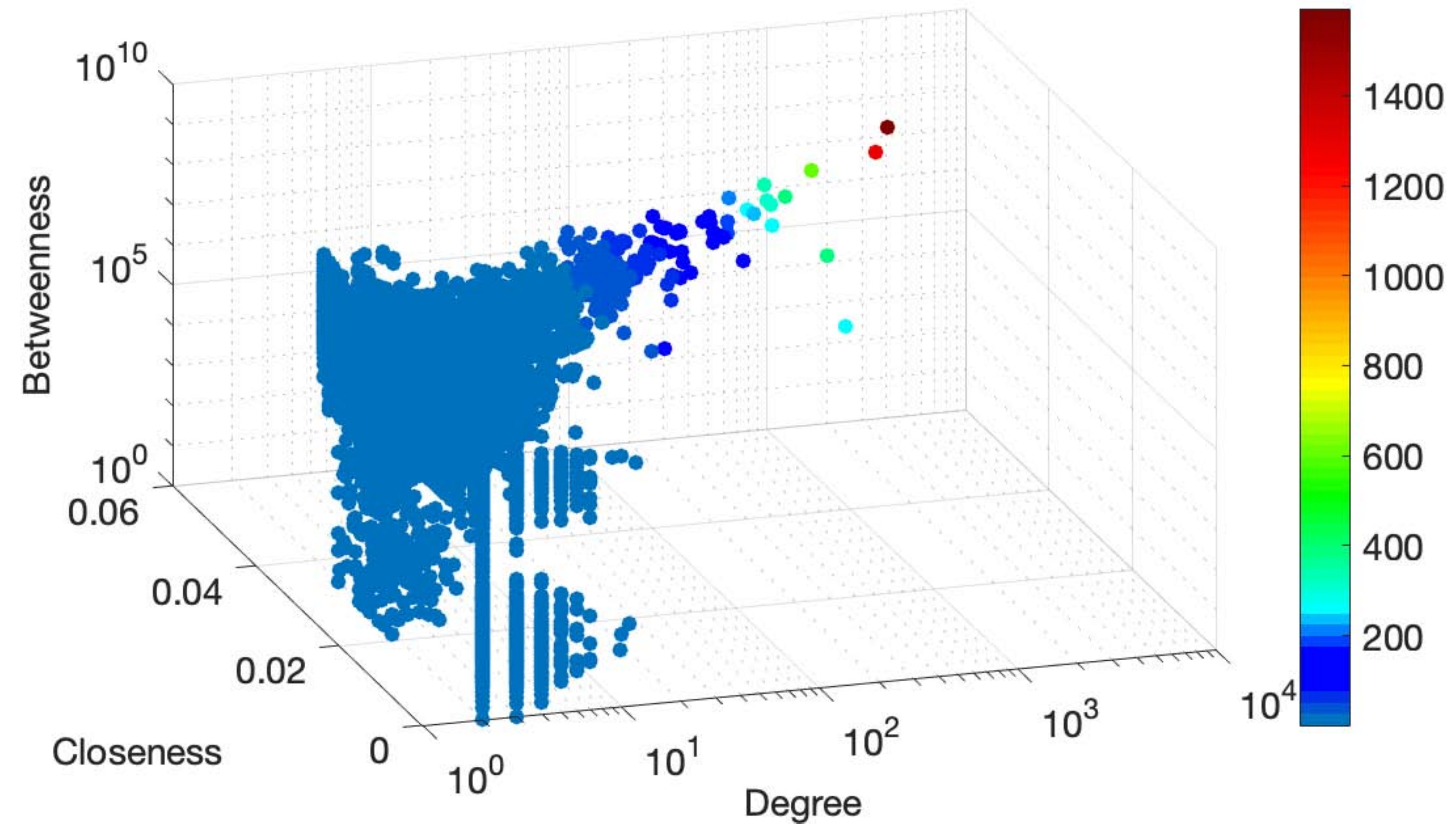}}
\end{subfloatrow}
\begin{subfloatrow}
 \ffigbox[\FBwidth][]
    {\caption{Average in-degree of the closest neighbors versus node out-degree.}\label{fig:kcn}}
    {\includegraphics[width=3.8cm]{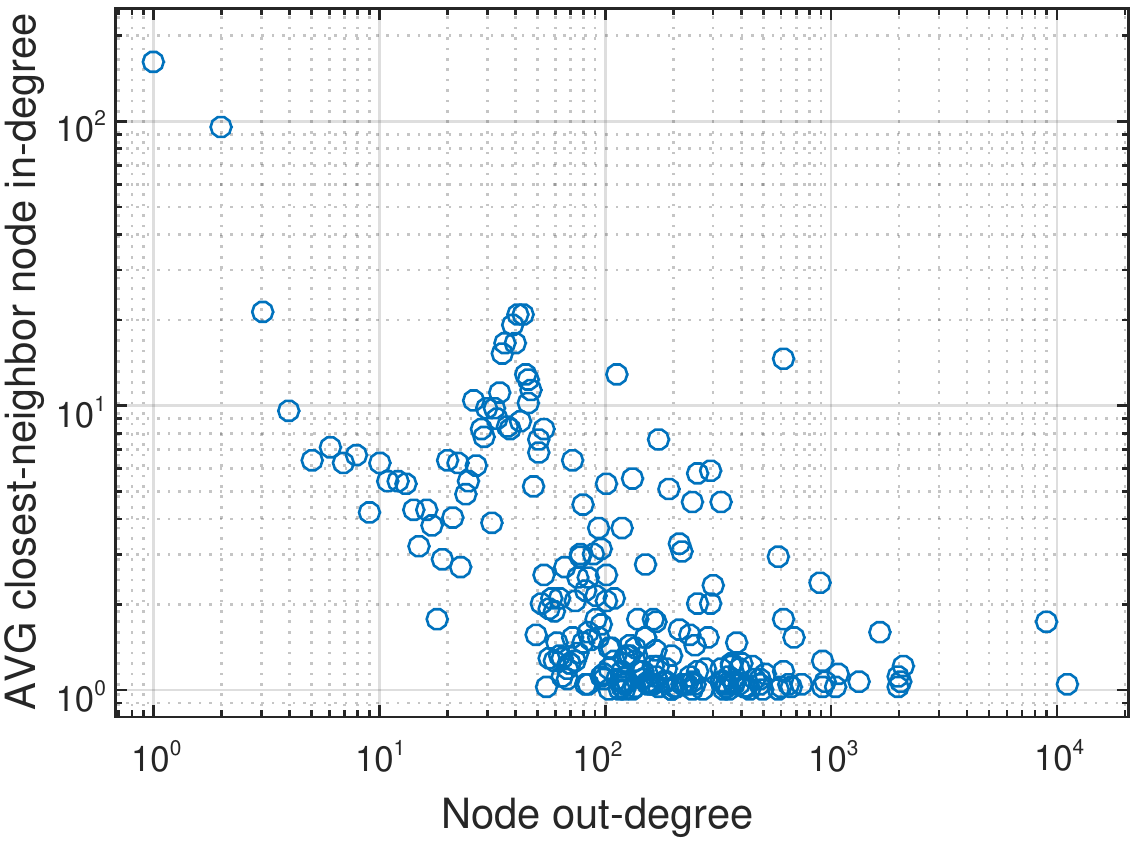}}
\end{subfloatrow}
}
{\caption{Centrality and Disassortativity of the Bitcoin network.}\label{fig:ck}}
\vspace*{-0.5cm}
\end{figure}

Furthermore, we introduce the \emph{betweenness centrality} denoted by $B(i)$ of node $i$ as follows,
\begin{equation}
\setlength{\abovedisplayskip}{3pt}
\footnotesize
B(i)=\sum_{u, v \in V} \frac{\sigma(u, v \mid i)}{\sigma(u, v)},
\setlength{\belowdisplayskip}{3pt}
\end{equation}
where $\sigma(u, v)$ is the amount of all existing shortest paths from node $u$ to node $v$, and $\sigma(u, v \mid i)$ is the number of those paths that meanwhile pass through node $i$. 

Fig.~\ref{fig:ck}(\subref{fig:centrality}) plots the closeness centrality and the betweenness centrality versus node degree. We observe from Fig.~\ref{fig:ck}(\subref{fig:centrality}) that the closeness increases with the increment of the degree. Meanwhile, there are a huge number of nodes with closeness values around 0.04, showing that they are relatively close to each other in the graph. These results further confirm that the Bitcoin network is a small-world network. On the other hand, only a few high-degree nodes have higher $O_{\text{WF}}(i)$ closeness values, indicating that they are tightly-connected with other nodes, i.e., conducting frequent transactions. From this observation, we speculate that those nodes may be some exchanges or related institutions, as mentioned in \S\ref{Connected component}. Fig.~\ref{fig:ck}(\subref{fig:centrality}) also shows that the betweenness increases with the increase of degree. Generally, if there are a large number of nodes with a high betweenness value, it will cause too many bridging nodes to appear in the graph. It makes the graph fragile, meaning that the network cannot maintain connectivity if some nodes are broken or leaving. These results imply that there are no such a lot of bridging nodes in the Bitcoin network. In other words, it is robust against node-removal.

We also observe that most nodes in the Bitcoin graph have relatively small values of both closeness and betweenness, implying that there are not too many central nodes. Only a few nodes have a relatively high closeness and betweenness. Thus, we observe a multi-central graph: some central nodes directly connect with a massive number of nodes without hops. The reason for the multi-center and excellent robustness can also be derived according to the \emph{disassortative distribution} of nodes. In the bitcoin network, the aforementioned disassortativity leads to a robust system, which will be discussed as follows.

\subsection{Disassortativity}\label{Disassortativity}

The vast gap between the number of high-degree nodes and that of low-degree nodes indicates the high \emph{heterogeneity} of the Bitcoin network. In order to further investigate the connection tendency of the Bitcoin network, we introduce the assortativity analysis. Particularly, we adopt the Pearson correlation coefficient denoted by $\rho$ to characterize the network assortativity~\cite{chen2014fundamentals}. The total number of edges in the graph is denoted by $M$. We then have Pearson correlation coefficient $\rho$ as follows,
\begin{equation}
\setlength{\abovedisplayskip}{3pt}
\scriptsize
\rho=\frac{M^{-1}\sum_{e_{ij}\in E(G)}^{}k_{i}k_{j}-[M^{-1}\sum_{e_{ij}\in E(G)}^{}1/2(k_{i}+k_{j})]^{2}}{M^{-1}\sum_{e_{ij}\in E(G)}^{}1/2(k_{i}^{2}+k_{j}^{2})-[M^{-1}\sum_{e_{ij}\in E(G)}^{}1/2(k_{i}+k_{j})]^{2}} ,
\setlength{\belowdisplayskip}{3pt}
\end{equation}
where $\displaystyle k_{i}$  is the out-degree of node $i$ at the beginning of link $\displaystyle e_{ij}\in E(G)$ and $\displaystyle k_{j}$ is the in-degree of the node at the end of link $\displaystyle e_{ij}\in E(G)$.  

In this case, the Pearson correlation coefficient is $\displaystyle \rho=-0.023$, indicating that the graph is disassortative. This is consistent with the results of~\cite{nerurkar2020dissecting}, which uses complete network data. However, the negative value of $\displaystyle \rho$ cannot fully indicate the \emph{disassortativity}. We then adopt the following measure $k^\text{cn-in}\left(k^\text{out}\right)$ to describe the average in-degree of the closest-neighbor nodes of node $i$ that has out-degree $k_{i}^\text{out}$,
\begin{equation}
\setlength{\abovedisplayskip}{3pt}
\footnotesize
k^{\text{cn-in}}\left(k^{\text{out}}\right)=\sum_{i=1, k_{i}^{\text{out}}=k^{\text{out}}}^{N}\left(\frac{k_{i}^{\text{cn-out}}}{N}\right) P(k^{\text{out}})  ,
\setlength{\belowdisplayskip}{3pt}
\end{equation}
where $k_{i}^\text{cn-out}=\sum_{j=1}^{N} a_{ij} k_{j}^\text{in} / k_{i}^\text{out}$, $\displaystyle a_{ij}$ is the $(i,j)$-th entry of the adjacency matrix in Eq.~\eqref{equ:aj}, and $\displaystyle P(k^\text{out})$ is the out-degree distribution function. If the value of $k^\text{cn-in}\left(k^{\text{out}}\right)$ shows a downward trend w.r.t. the variable $k^\text{out}$, then the graph is disassortative, as shown in the results in Fig.~\ref{fig:ck}(\subref{fig:kcn}). It means that high-degree nodes prefer connecting to low-degree nodes while low-degree nodes also prefer connecting to high-degree nodes. 
This effect can also be explained by the \emph{preferential attachment}, i.e., newly-joined nodes prefer connecting to high-degree nodes~\cite{kondor2014rich}.

\subsection{Rich-club coefficient}

The disassortativity implying that low-degree nodes tend to connect with high-degree nodes. Meanwhile, it is also necessary to explore the connectivity between high-degree nodes. In a complex network, the rich club refers to the phenomenon of a tight connection between high-degree nodes. In other words, the higher-degree nodes are regarded as the rich nodes, which nevertheless are more likely to gather into clubs (i.e., sub-graphs) comparing with those nodes with fewer edges. We denote the \emph{rich-club coefficient} by $\phi(k)$, which is defined as follows~\cite{colizza2006detecting},
\begin{equation}
\setlength{\abovedisplayskip}{3pt}
\footnotesize
\phi(k)=\frac{2 E_{>k}}{N_{>k}\left(N_{>k}-1\right)},
\setlength{\belowdisplayskip}{3pt}
\end{equation}
where $N_{>k}\left(N_{>k}-1\right) / 2$ is the maximum possible edges of all $N_{>k}$ nodes whose degree is higher than $k$; similarly, $E_{>k}$ denotes the number of edges among $N_{>k}$ nodes.

\begin{figure}
\floatsetup{captionskip=-0.001cm}
\ffigbox[8.9cm]{%
\begin{subfloatrow}
  \ffigbox[\FBwidth][]
   {\caption{Rich-club coefficient}\label{fig:rc1}}
    {\includegraphics[width=4.3cm]{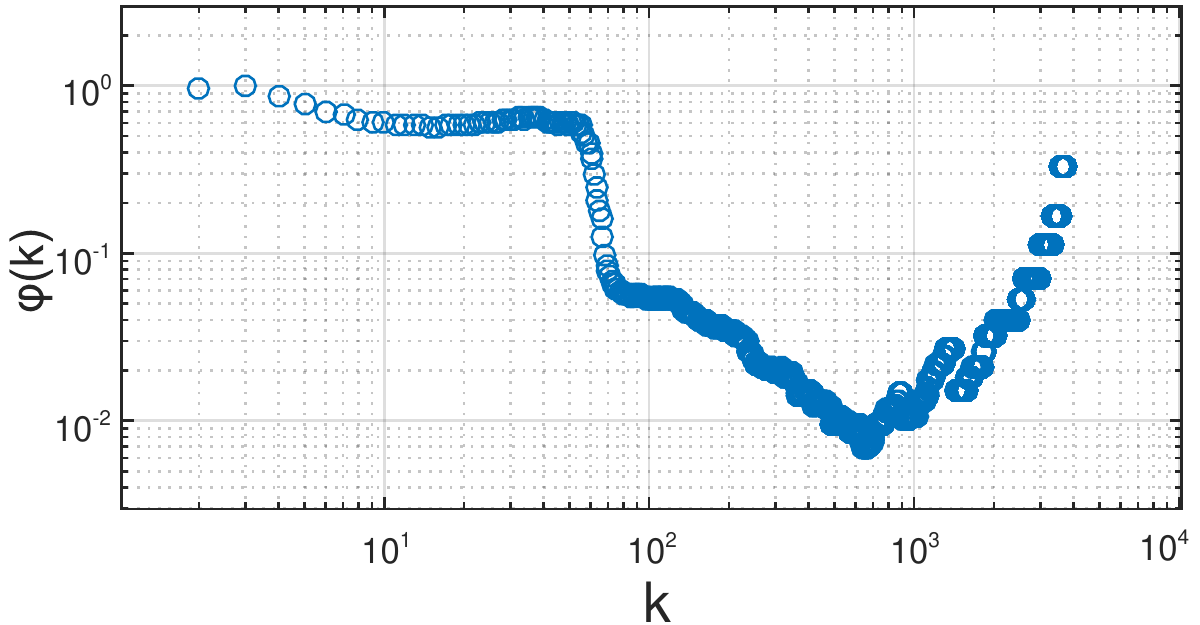}}
\end{subfloatrow}
\begin{subfloatrow}
 \ffigbox[\FBwidth][]
    {\caption{Normalized rich-club coefficient}\label{fig:rc2}}
    {\includegraphics[width=4.48cm]{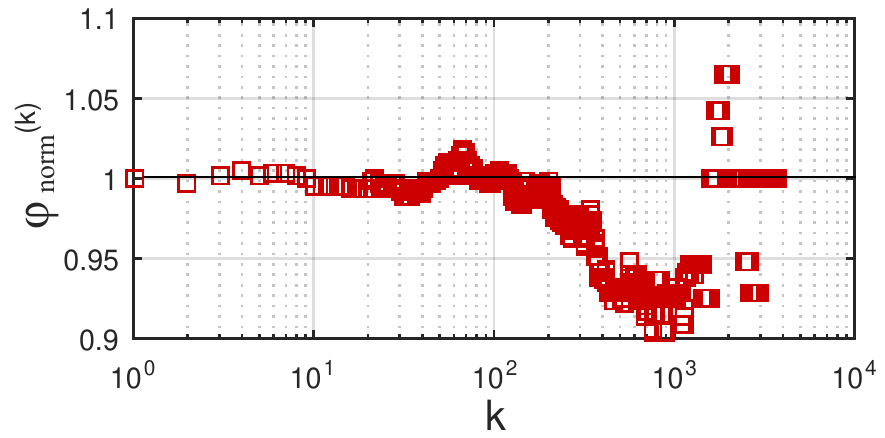}}
\end{subfloatrow}
}
{\caption{Non-rich-club effect of the network.}\label{fig:rc}}
\vspace*{-0.5cm}
\end{figure}

Fig.~\ref{fig:rc} plots the results. We observe from Fig.~\ref{fig:rc}(\subref{fig:rc1}) that the rich-club coefficient does not monotonically increase with the increment of $k$, implying no obvious rich-club phenomenon. 
For a more accurate evaluation, we adopt the normalized rich-club coefficient denoted by $\phi_{\text {norm }}(k)$ as follows,
\begin{equation}
\setlength{\abovedisplayskip}{3pt}
\footnotesize
\phi_{\text{norm }}(k)=\frac{\phi(k)}{\phi_{\text{rand }}(k)},
\setlength{\belowdisplayskip}{3pt}
\end{equation}
where $\phi(k)$ is the rich-club coefficient of the Bitcoin network and $\phi_{\text{rand }}(k)$ is the rich-club coefficient of a random network with the same degree distribution. Fig.~\ref{fig:rc}(\subref{fig:rc2}) plots the results, whereas the actual rich-club ordering depends on whether $\phi_{\text{norm }}(k)>1$. The results show an absence of rich-club ordering on most $k$ values.
 
In general, the Bitcoin network exhibits the non-rich-club phenomenon, implying that the high-degree central nodes in this network tend to disconnect with each other and are distributed in different connective sub-graphs. This effect can be explained by the fact that the central nodes are more likely to be some exchanges or large institutions with their own relatively-fixed customer groups. Therefore, the rich nodes are not closely connected to each other in the Bitcoin network.

\section{Conclusion}
In this brief, we have conducted a complex network analysis on the Bitcoin transaction network. In particular, we design a novel sampling method, namely random walk with flying-back properties, and obtain several important observations through analyzing sampled graphs. Firstly, the degree distribution of the Bitcoin transaction network conforms to a power-law distribution with the heavy-tail, approximated to a scale-free network. Secondly, we ascertain that the Bitcoin transaction network is a small-world network by analyzing the average clustering coefficient, the shortest-path length, and the small-world measurement. Thirdly, through the analysis of connected components, we find that most transactions are one-way deals. Moreover, we observe that the Bitcoin network is a multi-center robust network against the removal of nodes. Afterward, regarding the disassortativity of the Bitcoin network, low-degree nodes prefer connecting to nodes with higher degrees. We also find the preferential attachment of newly-added nodes. Finally, the current Bitcoin transaction network does not demonstrate the rich-club phenomenon. Such findings can help us better understand the structural behaviors of blockchain networks. Additionally, our analysis methods and sampling algorithms also provide insights in understanding networks with similar characteristics, such as scale-free networks, small-world networks and disassortative networks.

\bibliographystyle{ieeetr}  
\bibliography{ref}

\end{document}